\documentclass[manuscript]{acmart}

\AtBeginDocument{%
  \providecommand\BibTeX{{%
    \normalfont B\kern-0.5em{\scshape i\kern-0.25em b}\kern-0.8em\TeX}}}

\setcopyright{iw3c2w3}
\copyrightyear{2023}
\acmYear{2023}
\acmConference[ORSUM '23]{ORSUM '23: 6th Workshop on Online Recommender Systems and User Modeling, ACM RecSys 2023}{September 19th 2023}{Singapore}
\acmBooktitle{ORSUM '23: 6th Workshop on Online Recommender Systems and User Modeling, ACM RecSys 2023, September 19, 2023, Singapore}
\begin{document}

\title{TrueLearn: A Python Library for Personalised Informational Recommendations with (Implicit) Feedback}


\author{Yuxiang Qiu}\authornotemark[1]
\affiliation{%
 \institution{Department of Computer Science, University College London, UK}
}

\author{Karim Djemili}
\affiliation{%
 \institution{Department of Computer Science, University College London, UK}
}
\authornote{Authors contributed equally to the paper}
\author{Denis Elezi}\authornotemark[1]
\affiliation{%
 \institution{Department of Computer Science, University College London, UK}
}

\author{Aaneel Shalman}\authornotemark[1]
\affiliation{%
 \institution{Department of Computer Science, University College London, UK}
}

\author{Mar\'ia P\'erez-Ortiz}
\affiliation{%
 \institution{Centre for Artificial Intelligence, University College London, UK}
}

\author{Sahan Bulathwela}
\email{m.bulathwela@ucl.ac.uk}
\affiliation{%
 \institution{Centre for Artificial Intelligence, University College London, UK}
}
% \email{m.bulathwela@ucl.ac.uk}

% \author{Anonymous Authors}

% \affiliation{%
%  \institution{$^2\ $Jožef Stefan Institute, Jožef Stefan International Postgraduate School (Slovenia)}
% }

% \author{Mar\'ia P\'erez-Ortiz}
% \email{@ucl.ac.uk}
% \affiliation{%
%  \institution{University College London}
%  \streetaddress{Gower Street}
%  \city{London, UK}
%  \postcode{WC1E 6BT}
% }

%  \author{}
%  \email{@ucl.ac.uk}
%  \affiliation{%
%   \institution{University College London}
%   \streetaddress{Gower Street}
%   \city{London, UK}
%   \postcode{WC1E 6BT}
%  }
 
%   \author{John Shawe-Taylor}
%  \email{@ucl.ac.uk}
%  \affiliation{%
%   \institution{University College London}
%   \streetaddress{Gower Street}
%   \city{London, UK}
%   \postcode{WC1E 6BT}
%  }

\renewcommand{\shortauthors}{Bulathwela, et al.}

\begin{abstract}

This work describes the \emph{TrueLearn} Python library, which contains a family of online learning Bayesian models for building educational (or more generally, informational) recommendation systems. This family of models was designed following the "open learner" concept, using humanly-intuitive user representations. For the sake of interpretability and putting the user in control, the TrueLearn library also contains different representations to help end-users visualise the learner models, which may in the future facilitate user interaction with their own models.
Together with the library, we include a previously publicly released implicit feedback educational dataset with evaluation metrics to measure the performance of the models. The extensive documentation and coding examples make the library highly accessible to both machine learning developers and educational data mining and learning analytic practitioners.
The library and the support documentation with examples are available at \url{https://truelearn.readthedocs.io/en/latest}.

\end{abstract}

%%\begin{CCSXML}
%%<ccs2012>
%% <concept>
 %% <concept_id>10010520.10010553.10010562</concept_id>
  %%<concept_desc>Computer systems organization~Embedded systems</concept_desc>
  %%<concept_significance>500</concept_significance>
 %%</concept>
 %%<concept>
  %%<concept_id>10010520.10010575.10010755</concept_id>
  %%<concept_desc>Computer systems organization~Redundancy</concept_desc>
  %%<concept_significance>300</concept_significance>
 %%</concept>
 %%<concept>
  %%<concept_id>10010520.10010553.10010554</concept_id>
  %%<concept_desc>Computer systems organization~Robotics</concept_desc>
  %%<concept_significance>100</concept_significance>
 %%</concept>
 %%<concept>
  %%<concept_id>10003033.10003083.10003095</concept_id>
  %%<concept_desc>Networks~Network reliability</concept_desc>
  %%<concept_significance>100</concept_significance>
 %%</concept>
%%</ccs2012>
%%\end{CCSXML}
 \ccsdesc[500]{Information systems~Users and interactive retrieval}
 \ccsdesc[500]{Information systems~Personalization}
 \ccsdesc[500]{Information systems~Recommender systems}
% \ccsdesc[300]{Information systems~Search interfaces}
 \ccsdesc[100]{Information systems~Information extraction}
 \ccsdesc[500]{Applied computing~Interactive learning environments}

% \keywords{datasets, open education, engagement, video lectures, entity linking}

\maketitle

\section{Introduction}
It has been shown that personalised one-on-one learning could lead to improving learning gains by two standard deviations \cite{Bloom84}. 
With this goal in sight, and the ambition to democratise education to a world population, we require responsible Artificial Intelligence systems that can bring scalable, personalised and governable models to a mass of learners \cite{democratise2021}.
Up until recently, the go-to solution for scaling education has been Intelligent Tutoring Systems (ITS), heavily relying on testing users for knowledge, which is a practical option for formal courses with a limited number of learning materials involved. However, educational recommender systems have now the opportunity to go one step further, leveraging implicit interaction signals (such as clicks and watch time) to personalise and support learning for informal lifelong learners \cite{bulathwela2022sus}. This is exactly the focus of the models in this library, with the aim of making these methods more accessible, as publicly available learner models and datasets are currently scarce, and they can open up huge opportunities for education.

\subsection{Our Contribution} 

This work introduces \emph{TrueLearn}\footnote{Documentation available at \url{https://truelearn.readthedocs.io/en/latest}}, an open-source Python library that packages state-of-the-art online recommendation models, datasets and visualisation tools.
% Among a wide variety of use cases, it can be used e.g. to incorporate a personalisation component in an e-learning platform (e.g. YouTube and EdX) that can capture user clicks and watch time (of videos/lectures) by estimating the potential engagement of a learner with a learning resource, modelling relevant variables such as background knowledge, interest and novelty.
Among its diverse use cases, it can be used as a personalization component of e-learning platforms (e.g., YouTube and EdX) to estimate learners' potential engagement with learning resources and to model their background knowledge, interests, and novelty.
The library contains different components that will enable i) creating content representations of learning resources ii) managing user/learner states, iii) modelling the state evolution of learners using interactions and iv) evaluating engagement predictions. Requiring minimal data, its design offers a transparent solution that respects the privacy of its users and enables user interaction. The development of the TrueLearn library aims to provide both the research and developer communities with the opportunity to use the TrueLearn family of models.
The paper describes the development process and experiments that demonstrate the utility of this package to the educational data mining community and beyond.

While the motivation for TrueLearn stems from education, the models are applicable to a wide variety of applications that relate to informational recommendations and to model engagement in tasks in which human learning is involved. Additionally, note that the models included are suitable both for implicit and explicit feedback. In our experiments, we used a dataset of video lecture watch patterns, which we use as a proxy for learner engagement, but the same models could be applicable if learners also provided e.g. explicit feedback on the difficulty of the learning material.      

\section{Related Work} \label{lit_review}

We have researched related work on learner models and how to design usable machine learning libraries to make decisions regarding the design of the TrueLearn library. This section reviews these works and their influence on the development of the library. 

\subsection{Item Response Theory and Knowledge Tracing}

% As mentioned in \textit{A Family of Bayesian Algorithms to Match Lifelong Learners to Open Educational Resource} 
Item Response Theory (IRT) focuses on designing, analysing and scoring ability tests by modelling learner’s knowledge and question difficulty, without considering changes in knowledge over time. The simplest of IRT, the Rasch model \cite{Rasch1960}, computes the probability of scoring a correct answer as a function of the learner’s skill $\theta_l$ and the difficulty of the question/resource $d_r$:
\begin{equation}
P(correct\:answer|\theta_l, d_r) = f(\theta_l - d_r)
\end{equation}
where $f$ is usually a logistic function. TrueSkill model extends IRT to model the skill of multiple users playing a video game \cite{trueskill}. The TrueLearn models implemented in this work extend TrueSkill for learner engagement prediction.

% By replacing the learner and resource with two different players, we obtain the Elo rating system \cite{elo}, commonly used to rank chess players based on their game outcomes. The TrueSkill algorithm \cite{trueskill} takes this a step further, allowing entire teams of players to compete and adding a dynamic component to update player skills over time. 
% The original TrueLearn algorithms \cite{truelearn}, which model learning as a game between a team composed of the knowledge components (KCs) of the learner and an opposing team made up of the KCs covered by the learning resource, use TrueSkill to update the learner's skill based on the outcome of the match. TrueSkill serves as the foundation of TrueLearn and as such acts as a fundamental dependency in the library.

% \subsection{Knowledge Tracing (KT) and pyBKT}
An alternative to IRT for modelling learning is Knowledge Tracing (KT) \cite{corbett1994knowledge}. Unlike IRT, KT model does not consider question difficulty but instead estimates knowledge acquisition as a function of practice opportunities. Several Bayesian KT (BKT) algorithms that extend the original KT model have been proposed in the literature. The pyBKT library, a Python library of KT models provides a clear Application Programming Interface advocating the separation of data generation, model fitting and prediction functions \cite{pybkt}. However, conventional KT models do not train using online learning, introducing challenges when scaling to real-time scenarios with a large number of users learning over a long period of time. 
% While online formulations of KT exist \cite{bishopsnewbook}, they are not included in openly released code libraries.  
% , also ensures the models' correctness and computational efficiency through testing . pyBKT is of relevance to this project since, by bundling BKT algorithms into an open-source package, it makes them accessible to other researchers and developers, which is precisely what TrueLearn aims to do with the existing TrueLearn algorithms.

\subsection{TrueLearn Models}

The TrueLearn family of online Bayesian learner models uses implicit feedback from learners to recover their learning state.
% Several flavours of learners that model learners' interests, knowledge and novelty of materials are proposed in prior work. 
Prior work has proposed several learner models that capture the learner's interests and the knowledge and novelty of the material.
Subsequent work combines these individual models to propose proposed ensembles that can account for these factors simultaneously, improving the predictive performance of individual models \cite{bulathwela2022sus}. 

While being data efficient and privacy-preserving by design, TrueLearn models generate humanly intuitive learner representations that are inspired by open learner models. An open learner model is a learner model that has been made accessible to the learner it represents or to other users (e.g. teachers, parents) \cite{openlearnermodels}. This involves generating visualisations to construct an interface that will communicate to learners information about their knowledge and learning path. 
Open learner models come with definite advantages, such as promoting learner reflection by aiding learners in planning and monitoring their learning and allowing them to compare their knowledge to that of their peers \cite{openlearnermodels}. Open learner models also come with associated challenges, since all leaner representations and their visual presentations may not be equally understood by a wide variety of end-users. Among many different visualisations used to present learner knowledge state, user studies have shown that some visualisations are comparatively more user-friendly than others \cite{10.1007/978-3-319-98572-5_40,visualisationscomparison}. This work builds on these findings to develop a set of visualisations that aid this communication process.

% As such, it was of great importance to select the visualisations forming the interface appropriately by considering their effectiveness in communicating information about the learner model, and their ease of understanding to ensure that all learners, regardless of their background, would be able to benefit from them. Research ranking visualisations on these two criteria already existed () and we referred to it when determining which visualisations our library would be able to generate.

\subsection{Design a Machine Learning Library}

% if needed, this can be removed
% we only need to change the first sentence of the next paragraph
% While many design principles \cite{martin2000design} and patterns \cite{gamma1995design} have been proposed to address the problem of designing scalable and maintainable software, ...
To design a user-friendly, easy-to-use and scalable library, we need to avoid commonly used bad design practices, such as rigidity, fragility, immobility and viscosity \cite{martin2000design}. Rigidity refers to the tendency for software to be difficult to change. Fragility is the tendency for software to break once it has been updated. Immobility is the inability to reuse code within or across projects. Viscosity refers to the difficulty of retaining the original design when changes to the software are required. Various design principles and patterns are proposed and employed in software engineering to overcome these issues \cite{gamma1995design}. 
% For example, the strategy pattern makes code more flexible and reusable by deferring algorithm selection to runtime; the iterator pattern encourages the use of iterators to traverse and access data in containers, thus decoupling algorithms from data and resulting in easier software updates and code reuse .

When designing a machine learning library, we also need to account for unique challenges (e.g. incorporating data, pre-processing, models etc.). A great example of a well-designed machine learning library that has been taken up by both industry and academia recently is \texttt{scikit-learn}. 
% While the above patterns solve many of the problems of designing software, designing machine learning libraries presents unique challenges because it focuses more on models, data, parameters, and hyperparameters. This requires the library designers to consider how users set the hyperparameters of the model, prepare data for the training process, train the model based on the data, and use the trained model to accomplish their tasks. 
\texttt{scikit-learn} \cite{pedregosa2011scikit} proposes some general design principles (consistency, inspection and sensible defaults) and interface design (estimators and predictors) \cite{buitinck2013scikitapi} for building a scalable and user-friendly machine learning library. Consistency emphasises the importance of establishing a shared and consistent interface across different machine learning models, as this reduces the learning cost of the library. Inspection is concerned with exposing the model's parameters and hyperparameters as public attributes, which makes it easier for users to access the internal states of the model. Sensible defaults ensure that the model behaves reasonably well with the default values.
The estimator and predictor interfaces in scikit-learn reflect how the library implements these general guidelines. The estimator interface specifies a \texttt{fit} function to provide a consistent interface to the training model and exposes the \texttt{coef\_} attribute to facilitate the inspection of the internal state of the model. The predictor interface specifies the \texttt{predict} and \texttt{predict\_prob} functions as methods for utilising the trained model. Due to the time-tested design decisions that have succeeded in scikit-learn, the design decisions made in developing the TrueLearn library are inspired by these practices.

Because of the concept of duck typing in Python, others' model implementations can interoperate with scikit-learn (e.g., developers can plug them into scikit-learn's grid search) without being forced to inherit the above interfaces. This makes scikit-learn extensible and encourages users to reuse code. However, the use of duck typing makes it difficult to perform static program analysis \cite{milojkovic2017duck}, thus postponing the discovery of incorrect implementations until runtime and increasing the likelihood of software bugs \cite{chen2020typing}. Therefore, in our work, we tend to take a hybrid approach, utilising type annotations \cite{pep484} throughout the code base while allowing the use of static duck types supported by the \texttt{Protocol} class \cite{pep544}. Compared to traditional duck typing, static duck typing allows the library implementer to represent the requirement for parameters of a method explicitly but also does not force the user to inherit any class, making it easier for users to understand the intent of the method and helping static type checkers to analyse the code \cite{pep544}.

 \section{Library Overview}

This section describes the problem setting, the architecture of the library and how it can be applied in practice. While TrueLearn provides a probability that can be mapped to a binary outcome (engaged/not engaged), the probability prediction on different materials can rank them in relation to the state of the learner, creating personalised recommendations.

\subsection{Problem Setting}

The scenario educational recommendation focuses on is modelling a learner $\ell$ in learner population $L$ interacting with a series of educational resources $S_\ell \subset \{r_1, \ldots, r_{R}\}$ where $r_x$ are \emph{fragments/parts} of different educational videos. The watch interactions happen over a period of $T$ time steps, $R$ being the total number of resources in the system.
In this system with a total $N$ unique knowledge components (KCs), resource $r_x$ is characterised by a set of top KCs or topics $K_{r_x} \subset \{1, \ldots, N \}$. We assume the presence $i_{r_x}$ of KC in resource $r_x$ and the degree $d_{r_x}$ of KC coverage in the educational resource is observable.

% TODO: {-1, 1} or {0, 1}?
The key idea is to model the probability of engagement $e_{\ell, r_x}^{t} \in \{ 1, -1\}$ between learner $\ell$ and resource $r_x$ at time $t$ as a function of the learner interest $\theta^t_{\ell_{\texttt{I}}}$, knowledge $\theta^t_{\ell_{\texttt{NK}}}$ based on the top KCs covered $K_{r_x}$ using their presence $i_{r_x}$, and depth of topic coverage $d_{r_x}$.

% According to Bayes rule the posterior distributions are proportional to:
% \begin{equation} 
%  P(\theta^t_{\ell_{\texttt{I}}} | e^{t}_{\ell,r_x}, K_{r_x}, i_{r_x}) \propto P(e^{t}_{\ell,r_x} | \theta^t_{\ell_{\texttt{I}}}, K_{r_x}, i_{r_x}) \cdot P(\theta^t_{\ell_{\texttt{I}}})
% \end{equation}
% \begin{equation} 
%  P(\theta^t_{\ell_{\texttt{NK}}} | e^{t}_{\ell,r_x}, K_{r_x}, d_{r_x}) \propto P( e^{t}_{\ell,r_x} | \theta^t_{\ell_{\texttt{NK}}}, K_{r_x}, d_{r_x}) \cdot P(\theta^t_{\ell_{\texttt{NK}}})
% \end{equation}

\subsection{Architecture}

\begin{figure}[ht]
% \vskip 0.2in
\begin{center}
\centerline{\includegraphics[width=0.45\columnwidth]{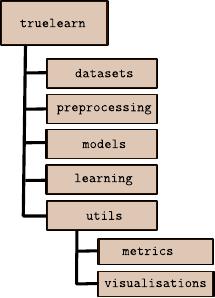}}
\caption{Library Structure of TrueLearn Library.}
\label{fig:structure}
\end{center}
% \vskip -0.2in
\end{figure}

The TrueLearn library consists of several modules that contain programming logic to execute different tasks using the library. Figure \ref{fig:structure} outlines the main structure of the TrueLearn library. These modules are described below.

\subsubsection{Datasets}
The TrueLearn dataset module integrates tools for both downloading and parsing learner engagement datasets. Currently, PEEK \cite{peek_orsum}, a publicly available learner engagement dataset is integrated. This module serves as a helper to integrate publicly available datasets that can be used for conducting experiments, evaluating model performance, and analysing learner data using TrueLearn’s visualisation capabilities.

\subsubsection{Pre-processing}
The pre-processing module contains utility classes designed specifically for extracting content representations from educational materials. The extracted representations serve as the foundation for creating KCs that can be used with IRT, KT and TrueLearn models. At present, utility functions that create Wikipedia-based KCs that are used in TrueLearn experiments \cite{truelearn} are included. 

\subsubsection{Models}
The \emph{models} module houses the class that can store the learner model. In this context, the learner model refers to the data structure that represents the learner's state (e.g. knowledge or interest). This learner model in the library is loosely coupled with the learning algorithms which makes this object reusable with many other learning algorithms that go beyond the TrueLearn algorithms currently included in the library. This means that the output of other learning algorithms such as BKT and IRT can still be used to create learner representations using this class.
% In the models module, the term “models” refers to the structured representations of a learner, their knowledge, and learning events. These abstractions are used to organise and store learning-related data but should not be confused with machine learning models. 

\subsubsection{Learning}
The learning module contains the implementation of TrueLearn algorithms that can perform training and prediction of learner engagement with transcribed videos \cite{trueeducation,bulathwela2022sus}. Each classifier within this module follows an implicit interface inspired by the scikit-learn design. For training,  \texttt{fit} function is used. For prediction, \texttt{predict} and \texttt{predict\_proba} functions are used to generate a binary label and a probability value respectively. Currently, i) a set of baseline models, ii) TrueLearn algorithms that model interest, novelty and knowledge in isolation, and iii) an ensemble model that combines the isolated models are implemented. 

\subsubsection{Metrics}
To evaluate the learning algorithm performance, the metrics module provides an interface to several key classification metrics including precision, accuracy, recall and F1 score. We use the scikit-learn API to support evaluation metrics. This opens up the opportunity to easily incorporate more evaluation metrics without having to put significant effort into testing and maintaining them in the future. 

\subsubsection{Visualisations}
To effectively depict the learner state, \emph{nine} different visualisations have been developed. These visualisations can be used with the ability to sort the output based on specific study topics (KCs), based on learners’ proficiency. Figure \ref{fig:vis} provides a preview of a subset of available visualisations. Out of these visualisations, seven are interactive visualisations that allow the end user to click and hover over the output to explore more details. However, they can also be saved as static images. The remaining two, namely, the Bubble Chart and Word Cloud, are exclusively static representations due to the limitations of the libraries used for their implementation. This module also provides the functionality to export these visualisations, where dynamic output can be saved in HTML format while the static output can be saved in various image formats such as PNG, JPEG and SVG. 

\subsection{Visualising the Learner State}

\begin{figure*}[ht]
% \vskip 0.2in
\begin{center}
\centerline{\includegraphics[width=1.1\linewidth]{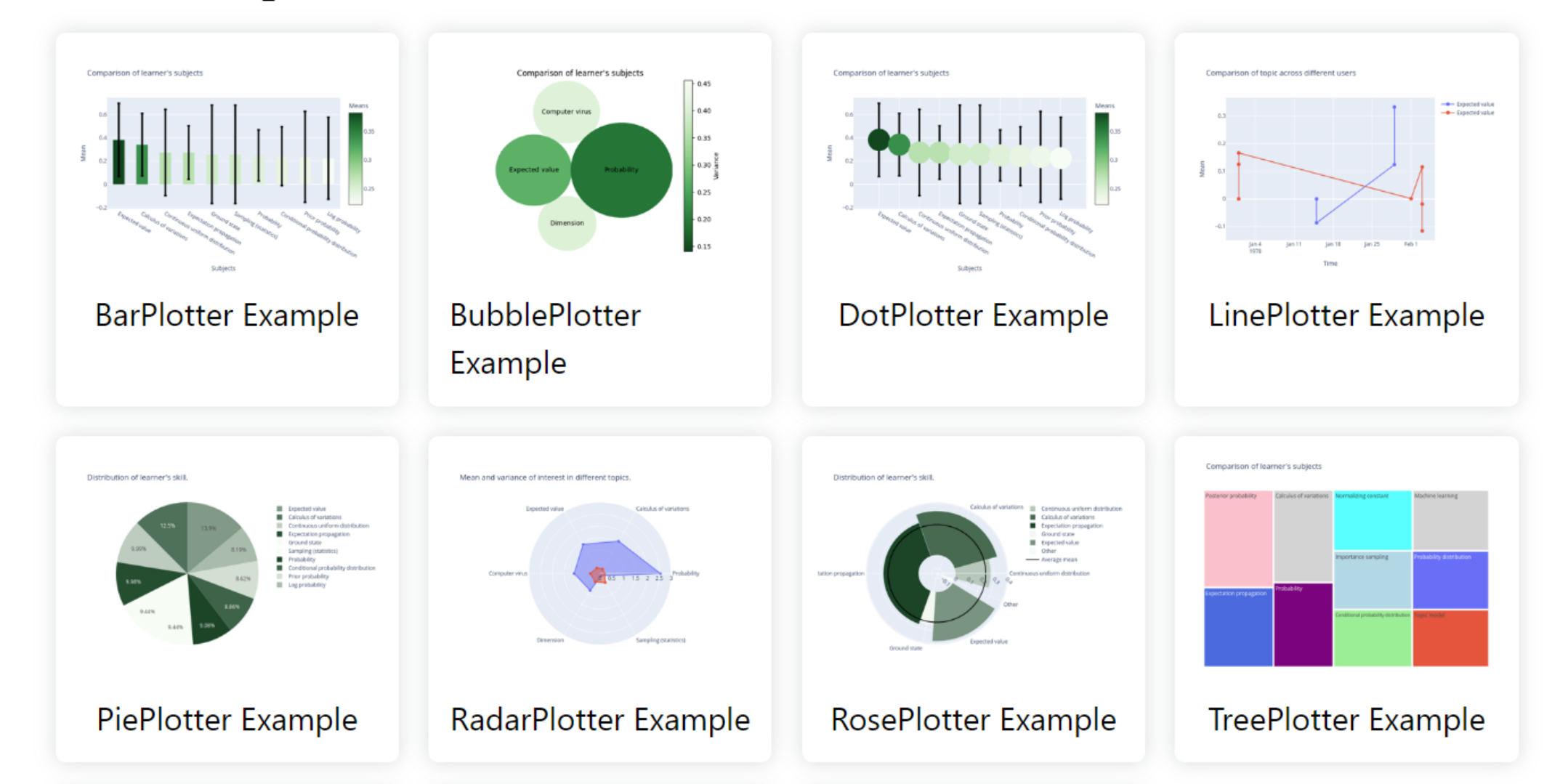}}
\caption{Multiple visualisations are available in TrueLearn to present the learner state in a humanly-intuitive way. Many different forms of visualisations are available to present the current learner state, while the line plotter also allows visualising the learner state's evolution over time.}
\label{fig:vis}
\end{center}
% \vskip -0.2in
\end{figure*}

Several visualisations that communicate the AI's learner state representation to a human user are implemented as part of the TrueLearn library. These visualisations are inspired by the open learner model concept where models are developed to maintain a humanly intuitive representation \cite{bull2013visualising}. Furthermore, the visualisations utilise user-friendly cues and conventions to minimise the learning curve for the human learner.  
The visualisations mainly represent the \emph{current state} of the learner while there are also visualisations that can depict how the state of the learner evolves over time (where the x-axis of the plot is time). The i) bar plot, ii) bubble plot, iii) dot plot, iv) pie plot, radar plot, rose plot and tree plot are implemented. Figure \ref{fig:vis} previews these visualisations. The TrueLearn family of algorithms represents a state using a mean and a variance value. In two-dimensional plots such as bar plots and dot plots, the mean is the y-axis and the confidence intervals mark the variance. In circle-based plots such as bubble plots and rose plots, the radius of the circle represents the mean. The intensity of the colour maps to the variance of the estimate (dark being low variance).

\section{Usage of TrueLearn Python Library}

\begin{figure*}[ht]
% \vskip 0.2in
\begin{center}
\centerline{\includegraphics[width=1.1\linewidth]{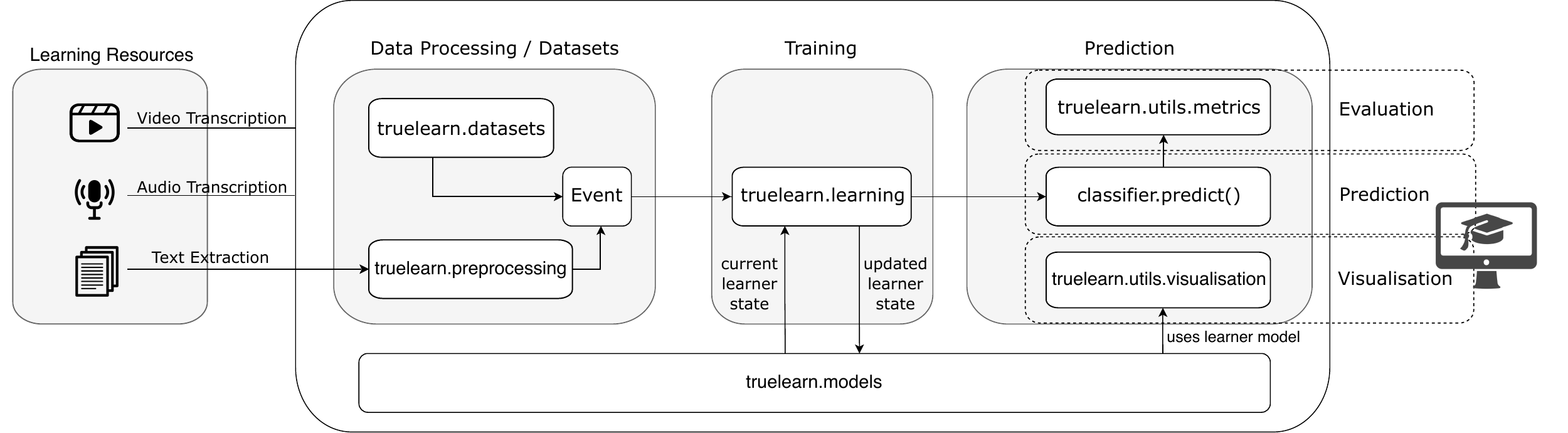}}
\caption{Potential usage scenarios of the TrueLearn Python package to build an educational recommender within an e-learning system.}
\label{fig:use}
\end{center}
% \vskip -0.2in
\end{figure*}

The TrueLearn Python library has two main uses, which are depicted in Figure \ref{fig:use}. 

% \begin{figure*}[ht]
% % \vskip 0.2in
% \begin{center}
% \centerline{\includegraphics[width=\linewidth]{system.pdf}}
% \caption{Potential usage scenarios of the TrueLearn Python package to build an educational recommender within an e-learning system.}
% \label{fig:use}
% \end{center}
% % \vskip -0.2in
% \end{figure*}

\subsection{Personalising E-learning/ Information}

This Python package makes it very easy to incorporate personalisation into video-based e-learning platforms. The pre-processing module allows extracting KCs/topics from text transcriptions of video-based learning materials by associating them with Wikipedia topics. When a learner starts watching a video, the interaction signals can be recorded from the web application. The TrueLearn package can instantiate a learner model for each individual learner in the platform and use the interaction logs to update the learner model. The online learning algorithms can continuously fit the events to the learner model. The updated model can be used to predict engagement with a set of potential future videos (or any other type of educational material) to rank them and provide back to the learner as recommendations. Additionally, the learner may request to see their current state at any given point. The visualisation module can be used with the current learner state to create both static and interactive learner-state visualisations that can be presented to the end-user.   

\subsection{Offline/Online Evaluation of Informational Recommenders}
Academics and researchers can use the TrueLearn library for conducting both online and offline evaluations of educational/informational recommendation algorithms. If the researchers need to benchmark a new learning algorithm, they can implement the learner algorithm using the common interface provided by the TrueLearn library. Then the Python library can be integrated with a web application to run online experiments and record user interactions. Similarly, offline evaluations can be done either by using i) an existing PEEK dataset or ii) integrating a new dataset. 

\section{Experiments}

In order to validate the accuracy of the implementation, we ran a few small-scale experiments attempting to replicate the results published in prior work \cite{bulathwela2022sus}. We used the PEEK dataset \cite{peek_orsum} to evaluate the performance of the primary TrueLearn models proposed in prior work, namely, i) TrueLearn Interest, TrueLearn Novelty, and TrueLearn INK. The experimental protocol was similar to the one used earlier.  
% We conducted a series of experiments using the PEEK dataset presented above to verify that the library was able to replicate the results obtained in the previous truelearn paper.
We also used a sequential experimental design. For each learner, its engagement at time $t$ is predicted using its engagement at times $1$ to $t-1$. We used the hold-out validation technique in our experiments where the training data is used for hyperparameter tuning. The best hyperparameter combination based on the F1-Score is identified. This combination is used with the test set to evaluate the final predictive performance. Since the engagement is predicted as a binary label in the PEEK dataset, the predictions for each event can be combined into a confusion matrix to compute accuracy, precision, recall, and F1 score. Same as in the prior publications, we calculate the weighted average of each learner's metrics based on their number of events.

% Q: How to reference the hyperparameters we used?
% \textbf{Hyperparameters}: For initial hyperparameters, we initialised the initial mean skill of learners to 0 for all models and use grid search to ﬁnd the suitable hyperparameters for the initial variance and beta value. 

\subsection{Empirical Evaluation}

The empirical results obtained are reported in Table \ref{tab:results}. The reported metric is the predictive performance each model obtained in the test set of the PEEK dataset.

% Q: How to properly reference previous results?
% \textbf{TrueLearn Novelty vs. TrueLearn Knowledge:} The experimental results show that TrueLearn Knowledge has the highest Acc. value, which validates the results in previous truelearn paper. In terms of recall and F1, TrueLearn Novelty outperforms TrueLearn Knowledge, which proves the need to exploit learner novelty and demonstrates that TrueLearn Novelty is indeed more competitive compared to TrueLearn Knowledge.

% \textbf{TrueLearn Interest vs. TrueLearn Novelty:} TrueLearn Novelty showed higher performance than TrueLearn Interest, as expected. Specifically, TrueLearn Novelty exhibited better performance than TrueLearn Interest in Acc., Prec., Rec. and F1.

% \section{Emperical Results}

% \begin{table}[ht]
% \caption{Predictive performance of TrueLearn Interest, TrueLearn Novelty, and TrueLearn INK algorithms are evaluated using Precision (Prec.), Recall (Rec.) and F1 Score (F1). The best and second best performance is indicated in \textbf{bold} and \emph{italic} faces respectively.}
% \label{tab:results}
% \begin{tabular}{ccccc}
% \toprule
% Model              & Acc. & Prec. & Rec. & F1    \\
% \midrule
% TrueLearn Interest &  58.04    & 57.88     & \textit{78.76}  & 62.95 \\ 
% TrueLearn Novelty  & \textit{64.78}	  & \textit{58.52}  & \textbf{80.91}  & \textbf{65.53} \\
% TrueLearn INK & \textbf{77.76}  & \textbf{64.29}  & 62.64 & \textit{63.23} \\
% \bottomrule
% \end{tabular}
% \end{table}

\begin{table}[ht]
\caption{Predictive performance of TrueLearn Interest, TrueLearn Novelty, and TrueLearn INK algorithms are evaluated using Precision (Prec.), Recall (Rec.) and F1 Score (F1). The best and second best performance is indicated in \textbf{bold} and \emph{italic} faces respectively.}
\label{tab:results}
\begin{tabular}{ccccc}
\toprule
Model              & Acc. & Prec. & Rec. & F1    \\
\midrule
TrueLearn Interest & 58.13    & 52.08     & \textit{78.61}  & 63.00 \\ 
% TrueLearn Novelty  & 65.65	  & 59.23	  & 79.81  & 65.71 \\
TrueLearn Novelty  & \textit{64.78}	  & \textit{58.52}  & \textbf{80.91}  & \textbf{65.53} \\
TrueLearn INK  & \textbf{78.32}  & \textbf{64.32}	  & {64.03}  & \textit{64.00} \\
\bottomrule
\end{tabular}
\end{table}

\subsection{Visualisations}
\begin{figure*}[ht]
% \vskip 0.2in
\begin{center}
\centerline{\includegraphics[width=.9\linewidth]{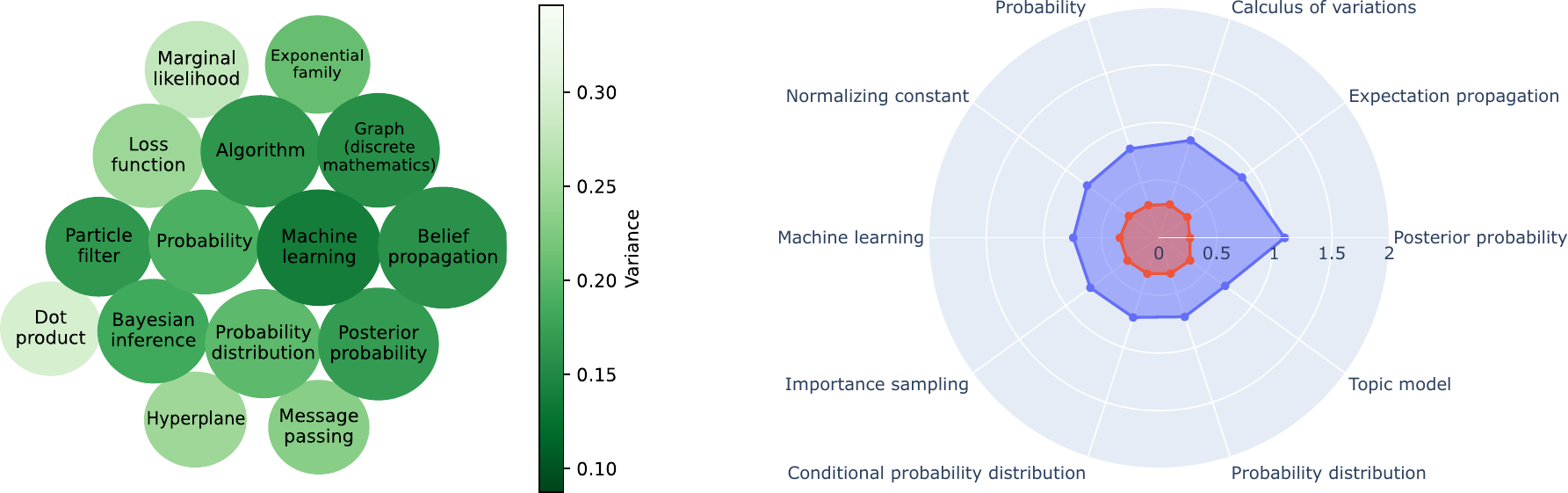}}
 \caption{Multiple visualisations are available in TrueLearn to present the learner state in a humanly-intuitive way. The top 15 (most knowledge acquired) topics/KCs of a learner in a (left) bubble plot representation, where the size of the circle aligns with the mean value of the knowledge parameter and the intensity of the colour maps to the variance (lighter for higher variance) and (right) radar plot representation, where the mean skill is red and the variance is blue.}
\label{fig:bubble}
\end{center}

% \vskip -0.2in
\end{figure*}

We designed visualisations, to empower students to recognise, contrast, and track the development of their skill level across subjects of their choice. It employs our algorithms’ prediction of their skill level (‘mean’) and calculation of a certain level in these predictions (‘variance’). These computations stem from learners' interaction with diverse learning resources and subject matters, classified based on Wikipedia topics. The library is designed to enable learners to effortlessly generate dynamic and static visualisations and potentially export static ones to provide learners with visually enriched insights, thereby promoting a learner-centric, self-regulated study experience. Figure \ref{fig:bubble} previews the learner knowledge state generated from one of the learners in the PEEK dataset test data. 

% properly cite these two papers
Our approach was guided by a thorough examination of seminal research on impactful learning visualisations. By incorporating the results from the questionnaire responses \cite{visualisationscomparison}, we arrived at a selection of nine distinct visualisations: Bar Charts, Line Charts, Dot Plots, Pie Charts, Rose Charts, Bubble Charts, Tree Maps, Radar Charts, and Word Clouds.

\section{Discussion}

We discuss the TrueLearn library using several functional and non-functional aspects of the TrueLearn Python library. 

\subsection{Library Design and Stability}

Throughout the development process, we adhere to design principles discussed in related work, such as consistency, inspection, and hybrid typing. By providing a consistent training and prediction interface for all classifiers in the TrueLearn library, we have achieved the consistency principle. Simultaneously, for easy inspection, we exposed the internal state of objects by means of property attributes, public attributes, and getters/setters. The reason why we use getters/setters instead of public attributes in the classifier is to better facilitate the hybrid typing approach. With these methods, we can perform type and value checks when the hyperparameters of the classifier are modified, which ensures the robustness of the classifier implementation and guides the user to pass in the correct hyperparameters. Considering that users may want to implement their version of the KC (in order to include custom topic fields needed by their subject domain), the TrueLearn classifier interacts with the KC by using static duck typing whenever possible, which promotes interoperability between TrueLearn and the client code as well as the scalability of TrueLearn itself.

% \subsection{Stability}

TrueLearn benefits from 100\% test coverage achieved through a combination of integration, unit, and documentation tests. These tests were run on Python versions ranging from 3.7 to 3.11 on all major operating systems to ensure compatibility. By making use of periodic testing with Continuous Integration, we can more greatly ensure that TrueLearn will work as intended regardless of the operating environment.

\subsection{Maintainability}

Modularity is an important aspect of the TrueLearn library, achieved by using a collection of Base classes that define a common interface and shared functionality. This design establishes a foundation that can be easily extended and customised moving forward.
To ensure the straightforward integration of the TrueLearn library, our documentation includes a wide range of examples showcasing the functionality offered by our API and visualisations. These examples are accompanied by the necessary code to generate them, allowing for a clear understanding of their implementation.
To further enhance maintainability, we have taken several additional measures. Firstly, we focused on minimising external dependencies wherever feasible to reduce the risk of compatibility issues and make it easier to maintain our codebase independently. Code consistency and readability are further enhanced by following the PEP 8 guidelines \cite{pep8}, which define a set of best practices for Python code. 

Furthermore, we provide a comprehensive API reference, offering detailed information for each class and function. The project website \footnote{Available at \url{https://truelearn.readthedocs.io/en/latest/}} describes relevant information for both a potential end-user or a contributor to familiarise with the library. Each of these comes with its own fine-grained examples along with descriptions of their purpose. To better explain the rationale of the technical and design decisions made, information is provided concerning the library's design and style guide in the \emph{contributing} section to minimise the learning curve associated with future development.

\subsection{Relevance and Impact}

{Modelling learner state in a humanly intuitive manner, requiring minimal data and exclusively relying on individual user actions, TrueLearn offers a transparent learner model that respects the privacy of its users and can scale to lifelong education. The development of the TrueLearn library aims to provide both the research and developer communities with the opportunity to seamlessly use the TrueLearn family of models in their work. The learner models utilise Wikipedia entity-based entity linking to create KCs that are based on a publicly available knowledge base. The content annotation also can scale to thousands of materials created in different modalities (video, text, audio etc.).}

{The impact of TrueLearn is two-fold. For development and research, the TrueLearn library employs a design that conforms with popular machine learning libraries such as \texttt{scikit-learn} and \texttt{pyBKT} \cite{buitinck2013scikitapi,pybkt}. The documentation is extensive and contains detailed examples that help the implementation. For end-use, the models employ probabilistic graphical models that are data efficient while providing humanly intuitive visualisations that trigger meta-cognition. The online learning algorithm updates the learner state in real-time helping better personalisation. A platform implementing TrueLearn can scale to a large population of users and support them through lifelong education.}

{The TrueLearn models advocate open learner modelling and employ open data sources such as Wikipedia helping the democratisation of education \cite{democratise2021}. The data and computational efficiency of the models also lead to minimised carbon footprint. The models are currently used in building a platform that connects Open Educational Resources to lifelong learners supporting lifelong, equitable education. In an era where AI is reaching a position that can impact society significantly, TrueLearn library unlocks the power of personalisation of information while taking into account multiple human values that go beyond knowledge management (e.g. climate responsibility, privacy, transparency to name a few).} 

\subsection{Limitations}

{While getting inspired by scikit-learn library, the learning algorithms in the TrueLearn library are not compatible with some helper functions available in scikit-learn (such as grid search). Building seamless compatibility with these utilities will enable the TrueLearn library to be adopted by a wider audience while minimising the development effort required to support such powerful features. The exclusive support of online learning algorithms can also be seen as a limitation of the current library as there are many batch learning algorithms that are proposed for educational recommendation \cite{intervention_bkt,ensemble_kt}. The library also doesn't support state-of-the-art deep learning algorithms \cite{deep_kt,pardos_serendipity}. These limitations are to be addressed in the future. }

% Karim: I wonder if this is better placed in Limitations?
% \textbf{Interoperability between truelearn and scikit-learn:} Currently, truelearn's classifier does not interoperate well with scikit-learn (e.g., using scikit-learn's grid search to find the right hyperparameter). The main problem comes from scikit-learn's requirement for the parameters of the `fit` function, which demands that the data for training the model (parameter X) be an array of shape \texttt{(n\_samples, n\_features)} where \texttt{n\_samples} denotes the number of samples and \texttt{n\_features} denotes the number of features. However, in the truelearn classifier, \texttt{fit} only takes in a single learning event. This makes it necessary for users to maintain a separate grid search algorithm when using truelearn, which increases the maintenance burden and the difficulty of reusing the code.

\section{Conclusion}
% \section{Summary of the System}

{This work showcases \textit{TrueLearn}, a Python library that models the learner knowledge and interest states to predict the engagement of learners with educational videos. The library contains several online learning models including ensemble models that model multiple factors that affect learner engagement together. It also includes a set of visualisations that can be used to interpret the learner's interest/knowledge state. The learner representations and state visualisations are comparable to outputs of knowledge tracing models, except, TrueLearn uses watch time interactions rather than relying on test taking. The empirical results demonstrate that the new implementation of the library achieves similar performance to the prior work that introduced these algorithms assuring correctness. The new implementation encourages educational data mining practitioners to use this library to incorporate educational video recommendations in e-learning systems. Researchers are encouraged to extend this library with new datasets and online learning algorithms for educational or informational recommendations.  }

\subsection{Future Work}

The immediate future work entails running several user studies to evaluate the effectiveness of the visualisations and identifying ways to improve them. Incorporating the library into a real-world e-learning platform to run online evaluations is also a top priority. We also aim to evaluate the performance of the TrueLearn models in the context of information recommenders that present informational content such as news and podcasts to demonstrate the generalisability of the models. Incorporating models that can exploit explicit feedback simultaneous to implicit feedback can also enhance the library and its utility. In the long term, we aim to add more general informational recommendation algorithms to the library and mobilise the research community to contribute various models, pre-processing techniques and evaluation metrics that the library can benefit from.

\begin{acks}
This work is partially supported by the European Commission-funded project "Humane AI: Toward AI Systems That Augment and Empower Humans by Understanding Us, our Society and the World Around Us" (grant 820437) and the X5GON project funded from the EU's Horizon 2020 research programme grant No 761758.
\end{acks}

\bibliographystyle{ACM-Reference-Format}
\bibliography{sample-base}

\end{document}